\documentclass[10pt]{article}
\usepackage{amssymb, amsmath, verbatim, epsfig,setspace,mathtools}

\usepackage{mathtools}
\newcommand{\fig}[1]{{\bf Fig.~\ref{#1}}}

\usepackage{xcolor}
\usepackage{amsmath,amssymb,graphicx,epstopdf,bm,bbm}

\usepackage{lscape}
\usepackage{graphicx}
\usepackage{multirow}
\usepackage[super,compress]{natbib}
\usepackage{siunitx,booktabs}

\usepackage{lipsum}  
\title{\bf 
\vspace{-0.6in}
\Large{Hair is a functionally graded composite, not a uniform fiber}
}

\date{\vspace{-3ex}} 
\author{Andrew K. Schulz$^{1,*}$, Svetlana Griego$^{1,2,3,+}$, Vesna Srot$^{4,+}$, Giulia Ballardini$^1$,\\Natalia Gonzalez-Vazquez$^{2,3}$, Devin Sheehan$^5$, Felicitas Predel$^{4}$, Cem Balda Dayan$^{6}$,\\Deepti S. Philip$^{1,2,3}$, Helena Vrencev$^5$, Jelena Lazovic$^5$, Peter A. van Aken$^{4}$,\\Gunther Richter$^{2,3}$, Katherine J. Kuchenbecker$^{1,3,*}$\\
\small{$^{1}$Haptic Intelligence Department, Max Planck Institute for Intelligent Systems (MPI-IS), Stuttgart, Germany,}\\
\small{$^{2}$Materials Central Scientific Facility, MPI-IS, Stuttgart, Germany,}\\
\text{\small{$^3$University of Stuttgart, Stuttgart, Germany,}}\\
\text{\small{$^4$Stuttgart Center for Electron Microscopy, Max Planck Institute for Solid State Research, Stuttgart, Germany,}}\\
\text{\small{$^{5}$Medical Systems Central Scientific Facility, MPI-IS, Stuttgart, Germany,}}\\
\text{\small{$^{6}$Robotic Materials Department, MPI-IS, Stuttgart, Germany}}
}

\usepackage[left=1.5cm,right=1.5cm,top=2cm,bottom=2cm]{geometry}
\usepackage{hyperref}

\begin{document}
\renewcommand{\figurename}{\textbf{Fig.}}
\newcommand{\pl}{\textcolor{white}{--}}
\def\urlprefix{}
\newcommand{\lw}{\textcolor{white}{$<$}}
\maketitle
\noindent \text{\small{+ These authors contributed equally, and names are listed alphabetically.}}\\ 
{\bf * Corresponding authors:}\\
Andrew K. Schulz and Katherine J. Kuchenbecker\\
Heisenbergstraße 3, 70569 Stuttgart, Germany\\
aschulz@is.mpg.de \& kjk@is.mpg.de

\doublespacing  
\begin{abstract} 
Hair provides mammals with diverse benefits, including protection~\cite{beseris_impact_2020,cohen_biological_2025}, thermoregulation~\cite{wu_biomimetic_2023,carolan_anti-icing_2025}, and enhanced sensory perception~\cite{kim_mechanism_2012,deiringer_functional_2023,takatoh_vibrissa_2018}. Unlike tendons and teeth, which are biomineralized~\cite{ping_mineralization_2022}, hair is hypothesized to accomplish its structure-function relationship purely through keratin~\cite{yang_strength_2020}, a fibrous protein that provides structural integrity~\cite{wang_keratin_2016}. Recent research showed that mechanical properties can vary substantially both within and across hair types: the whiskers of Asian elephant (\textit{Elephas maximus}) exhibit a two-order-of-magnitude material stiffness reduction from base to tip, whereas elephant body hairs are nearly homogenous~\cite{schulz2026functional}. Here, we demonstrate that three hierarchical structures vary significantly along the body hairs and whiskers of domestic cat (\textit{Felis catus}): the layered outer keratin wall, 250-nm-diameter melanosome-like~\cite{morioka_guide_2009,coroaba_new_2020} granules in the cortex, and calcium enrichment of these granules. As occasionally described for human hair~\cite{gorniak2014nano,itou2018morphological}, the oblong granules are arranged in longitudinal channels, potentially reinforcing the cortex; their prevalence correlates with local mechanical properties along the hair's length. Prolonged chemical treatment of body hairs removes calcium from the granules while breaking down the outer cuticle and internal cortex, hardening bases and splitting tips. Though previously assumed uniform, morphology, composition, and elemental enrichment can change along hairs, producing composite structures with functional gradients.
\end{abstract}

\section*{Main}  
Keratin is one of the most abundant biopolymers on Earth~\cite{rajabi_keratinous_2020}, creating diverse structures (e.g., nails, horns, hooves, hairs) that endow animals with a plethora of functional benefits~\cite{wang_keratin_2016}. Hair adorns all mammalian species and has been assumed to have a homogeneous composition of keratin since the 1960s~\cite{wang_keratin_2016,pautard_mineralization_1963}. The lipids and melanosomes (pigment granules) contained in hair~\cite{koch2020biology} have long been assumed not to affect the composite's functional properties~\cite{fernandes2023hair}. Recent research on whiskers, a type of sensory hair present on many mammals but not on humans, showed that both Asian elephants and domestic cats have whiskers that shift from a stiff material similar to rigid plastic at the base to a soft elastomer at the tip~\cite{schulz2026functional}, creating a functional gradient of stiffness spanning two orders of magnitude along the whisker length. Biological structures typically achieve such remarkable stiffness gradients either with fibers that shift orientation~\cite{alderete_does_2025,schulz_second_2025} or by including elements in their mineral phase~\cite{ping_mineralization_2022,badar_nonlinear_2025}, a process known as biomineralization. 

Biomineralization provides diverse functional benefits to biological structures~\cite{eder_biological_2018,arias_polysaccharides_2008}; for example, hydroxyapatite gives shrimp claws high hardness~\cite{chua_biomineralization_2023}, and calcium phosphates make bird and mammal tendons strong~\cite{zou_three-dimensional_2020}. Aside from whale baleen~\cite{szewciw_calcification_2010}, keratin's micro-scale organization and biomineralization status have rarely been reported, as investigations primarily focused on macroscopic morphology and mechanics~\cite{wang_keratin_2016}. Understanding how keratin-based biomaterials can exhibit multi-order-of-magnitude stiffness gradients remains an open question~\cite{schulz2026functional}. It seems likely that the arrangement of the keratin tunes the local material properties to produce the observed gradients.

We hypothesize that the material differences previously observed in whiskers and body hairs stem from variations in their local composition. We thus analyzed the bases and tips of domestic cat (\textit{Felis catus}) body hairs and whiskers (\fig{fig:calciumintro}a), observing their morphology, mechanics, and elemental composition across length scales ranging from millimeters ($10^{-3}$\,m) to nanometers ($10^{-9}$\,m) (\fig{fig:calciumintro}b). The results of our experiments demonstrate that the hair cortex contains calcium-enriched granules (CEGs) with varying prevalence and calcium enrichment levels that correlate with mechanical longitudinal gradients. These gradients differ between body hairs and whiskers and from the base to the tip of each hair type, accompanied by consistent thickening of the layered cuticle (\fig{fig:calciumintro}c--d). Our findings of substantial calcium enrichment of oblong granules organized in longitudinal channels within hairs showcase how biological composite materials can gain some of the mechanical benefits of biomineralization while keeping the compliance afforded to keratin.

\subsection*{Hair cortex contains longitudinal channels of calcium-enriched granules}
\begin{figure}[thbp]
      \centering
      \includegraphics[width=1\textwidth,page=1,trim=0.0in 0in 0in 0.0in,clip]{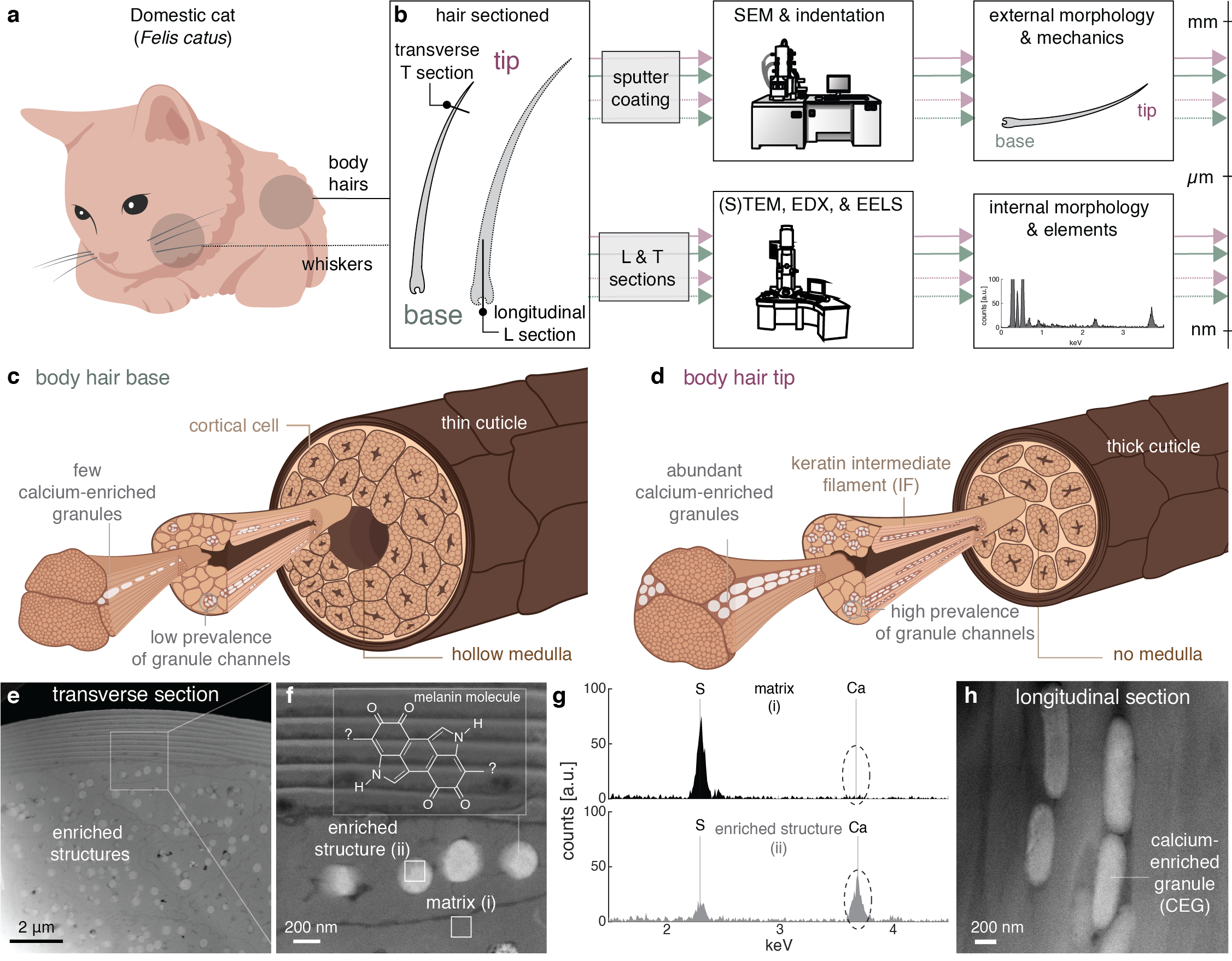}
      \caption{
      \textbf{$\mid$ Multi-length-scale characterization reveals calcium-enriched granules in hair cortex.} \textbf{a,} All experiments examined the bases and tips of body hairs (fur) and whiskers from domestic cat (\textit{Felix catus}). \textbf{b,} Experimental techniques used to examine the morphology, mechanics, and elements of specimens across different length scales. \textbf{c,} Illustration of the base structure of body hair showing the structural hierarchy, including a thin cuticle and an enlarged view of an intermediate filament bundle in a cortical cell with few calcium-enriched granule channels. \textbf{d,} Illustration of the tip structure of body hair, showing a thick cuticle, no medulla, and abundant granule channels. \textbf{e,} Transverse section high-angle annular dark-field scanning transmission electron microscopy (HAADF-STEM) image of untreated cat body hair showing the presence of enriched structures. \textbf{f,} Enlarged HAADF-STEM image with boxes labeling (i) a region of the matrix of the cortex and (ii) a calcium-enriched structure, overlaid with the chemical diagram of the molecular structure of melanin~\cite{morioka_guide_2009,coroaba_new_2020}. \textbf{g,} Energy-dispersive X-ray (EDX) spectra showing the sulfur and calcium content of the regions of the matrix and enriched structure marked in \textbf{f}. \textbf{h,} HAADF-STEM image of a longitudinal section of body hair displaying a side view of the enriched structures; their composition and melanosome-like appearance inspired the name calcium-enriched granule (CEG). 
      }
      \label{fig:calciumintro}
\end{figure}

Analysis of high-angle annular dark-field (HAADF) scanning transmission electron microscopy (STEM) images of transverse cross sections of domestic cat body hairs (fur) (\fig{fig:calciumintro}e) revealed a large number of heavier-element-enriched round structures with an average diameter of about 250\,nm (range of 200--400\,nm) inside the cortex (\fig{fig:calciumintro}f). Energy-dispersive X-ray (EDX) spectroscopy measurements showed these regions have less sulfur and more calcium than the surrounding cortex matrix (\fig{fig:calciumintro}g). Subsequent analysis of longitudinal cross sections of body hairs revealed the structure, morphological organization, and distribution of the enriched regions (\fig{fig:calciumintro}h). These oblong inclusions extend along the longitudinal axis of the hair with a length-to-diameter ratio ranging from about three to six. They share morphological characteristics with the pigment granules and melanosomes previously described in human hair~\cite{birbeck1956structure,koch2019variation}, yet their systematic organization into longitudinal channels has rarely been noted. We refer to these inclusions as calcium-enriched granules (CEGs) to highlight their elemental composition and similarity to human-hair structures~\cite{gorniak2014nano,itou2018morphological}. Pigment granules contain melanin, which is known to bond with calcium in other biomaterials~\cite{bush2007quantification,hong2007current,hoogduijn2003melanin}; furthermore, calcium has been found in the cortex of human hair and was tentatively attributed to melanosomes~\cite{merigoux2003evidence}. However, the prevalence and level of calcium enrichment of these granules have not been studied at different locations along the hair.

Individual granules and their surrounding matrix were investigated using electron energy loss spectroscopy (EELS) to understand bonding states~\cite{spath_microspectroscopic_2015} inside the cortex of hair fibers (\textcolor{violet}{Extended Data Fig.~1}a). EELS showed characteristic calcium L$_{2,3}$ edge peaks (\textcolor{violet}{Extended Data Fig.~1}b-c) for the CEGs; these peaks are absent in the matrix at both the base and tip. When comparing the fine structural details of C-K ELNES (\textcolor{violet}{Extended Data Fig.~1}c), we found base and tip CEGs display identical features that are distinct from the matrix at the same locations. The low-loss EELS spectra follow this same trend (\textcolor{violet}{Extended Data Fig.~1}d). Thus, the carbon bonding of the CEG material is different from that of the matrix, consistent with a distinct organic composition, potentially reflecting the enrichment of calcium-binding biomolecules (e.g., melanin~\cite{bush2007quantification,hong2007current}) or proteins (e.g., S100A3~\cite{halawi2014s100,unno2011refined}) within CEGs. The calcium in the CEGs might be bonded between or directly to the melanin monomers within the melanosomes of the hair fiber, which have open bonding sites~\cite{morioka_guide_2009,coroaba_new_2020} (\fig{fig:calciumintro}f), but further chemical experiments would be required to identify how calcium is bonded in cat body hairs. Therefore, we denote hair as a calcium-enriched biopolymer, not biomineralized.

Both body hairs and whiskers have long been assumed to consist of 99\% keratin intermediate filaments (IFs) arranged longitudinally along the hair and wrapped in a layered cuticle wall~\cite{yang_strength_2020}. Since mechanical properties can change dramatically along the length of a single keratin-based structure~\cite{schulz2026functional}, we hypothesize that body hairs and whiskers are biological composites that combine a keratinous wall, keratin IFs, and pigment granules with localized calcium enrichment, and that granule abundance and calcium-enrichment level vary both across hair types (body hairs versus whiskers) and along the hair length (base versus tip). We propose that this combination of keratin, granules, and calcium enrichment can produce diverse functional gradients that enable two biologically critical features: (1) mechanical reinforcement, as shown for the biological composites of tendons~\cite{ping_mineralization_2022}, bone~\cite{fratzl_natures_2007}, and whale baleen~\cite{szewciw_calcification_2010}; and (2) sensory benefits, as shown for the encoding of contact location by whiskers with functionally graded material stiffness~\cite{schulz2026functional}. 

\subsection*{Cuticle thickness, granule prevalence, and calcium enrichment vary longitudinally}
\begin{figure}[thbp]
      \centering
      \includegraphics[width=0.95\textwidth,page=1,trim=0.0in 0in 0in 0.0in,clip]{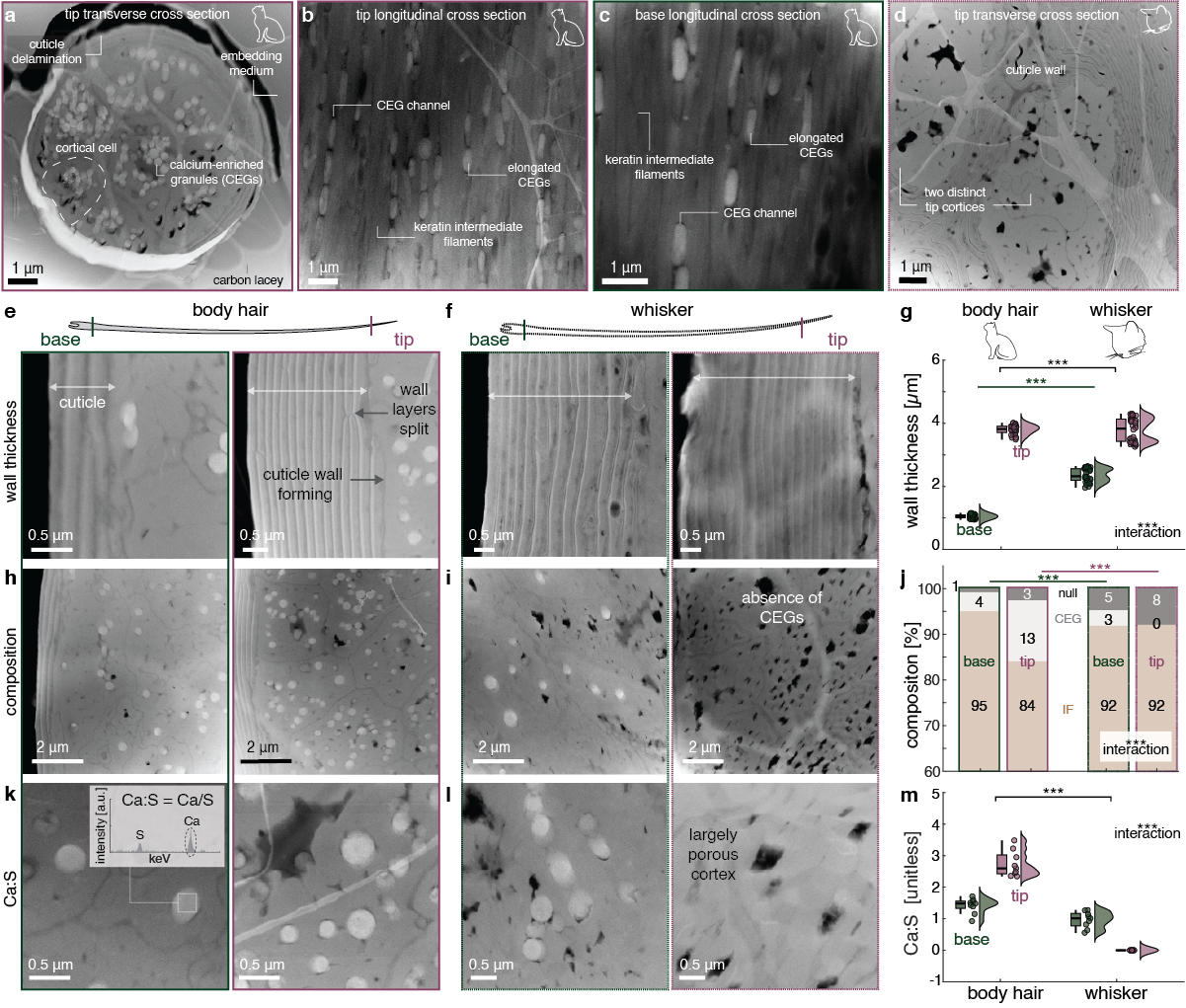}
      \caption{
      \textbf{$\mid$ Compositional gradients of cat body hairs and whiskers.} \textbf{a,} STEM image of a body hair tip, showing the complete transverse cross section with a diameter of about 8\,$\mu$m. \textbf{b,} STEM image of the longitudinal section of a body hair tip showing high prevalence of CEGs. \textbf{c,} STEM image of the longitudinal section of a body hair base showing fewer CEGs. \textbf{d,} STEM image of the transverse section of a cat whisker tip showing more than 20 layers of cuticle wall, two distinct cortex regions, and no CEGs. \textbf{e,} STEM images of body hair base and tip, showing the cuticle greatly increases in thickness longitudinally. \textbf{f,} STEM images of whisker base and tip, showing a moderate wall thickness increase. The cuticles in panels \textbf{e} and \textbf{f} have partially delaminated from the embedding medium, so only the inner part of the cuticle is visible, as also seen in panel \textbf{a}. \textbf{g,} Comparison of the cuticle wall thickness of body hair base ($n=40$) and tip ($n=40$) versus whisker base ($n=40$) and tip ($n=40$). The line represents the median, the dots show the raw measurements, and the violin illustrates their distribution. \textbf{h,} STEM images of regions of interest (ROIs) of the body hair base and tip cortex from which granule prevalence was determined. \textbf{i,} STEM images of ROIs of whisker base and tip, showing an absence of granules in the whisker tip's cortex. \textbf{j,} Comparison of mean measured cortex composition for body hair base ($n=20$ ROIs) and tip ($n=20$) versus whisker base ($n=20$) and tip ($n=20$); keratin intermediate filament (IF), calcium-enriched granules (CEGs), and hollow (null) percentages are provided for each specimen type. \textbf{k,} Enlarged STEM images of body hair base and tip, along with an inset of EDX spectra for the calcium-to-sulfur ratio (Ca:S) of a CEG. \textbf{l,} Enlarged STEM images of whisker base and tip cortex showing the presence and absence of CEGs, respectively. \textbf{m,} Comparisons of Ca:S ratio across body hair base ($n=9$) and tip ($n=9$) versus whisker base ($n=9$) and tip ($n=9$, taken from the matrix surrounding voids). Statistically significant differences in panels \textbf{g}, \textbf{j}, and \textbf{m} are shown as: $^{\star\star\star}~p < 0.001$. All STEM images in this figure are HAADF-STEM images.  
      }
      \label{fig:TEM_raw_whiskers}
\end{figure}

Additional STEM imaging of transverse (\fig{fig:TEM_raw_whiskers}a) and longitudinal sections (\fig{fig:TEM_raw_whiskers}b--c) of cat body hairs and whiskers revealed several morphological differences between the two hair types and from the base to the tip. Transverse scanning electron microscopy (SEM) and STEM images along the length of body hairs (\textcolor{violet}{Extended Data Fig.~2}a--d) show that the CEGs are sporadically distributed throughout the cortex and never occur in the cuticle (\fig{fig:calciumintro}e,h, \ref{fig:TEM_raw_whiskers}a--c), with more CEGs visible at the tip than the base. By contrast, the cat whisker's composition changes in the opposite direction, with no granules, with or without calcium enrichment, found at the tip. STEM analysis shows that the cuticle accounts for about 88\% of the cross-sectional area of the sectioned cat whisker tip (\textcolor{violet}{Extended Data Fig.~3}), and one sample contains two distinct cortex regions (\fig{fig:TEM_raw_whiskers}d).

The cuticle of the cat body hair (\fig{fig:TEM_raw_whiskers}e) increases in thickness by almost 300\% from the base ($1.0 \pm 0.06$\,$\mu$m) to the tip ($3.8 \pm 0.13$\,$\mu$m), contradicting previous assumptions of uniform cuticle thickness along hairs~\cite{breakspear_cuticle_2022,voges_structural_2012}. The cuticle thickness of the cat whisker (\fig{fig:TEM_raw_whiskers}f) increases by almost 60\% from the base ($2.4 \pm 0.2$\,$\mu$m) to the tip ($3.8 \pm 0.4$\,$\mu$m). Statistical analysis of these changes (\fig{fig:TEM_raw_whiskers}g) revealed significant effects of hair type ($F(1,78) = 363.17$, $p < 0.001$), location ($F(1,78) = 3020.99$, $p < 0.001$), and their interaction ($F(1,78) = 307.45$, $p < 0.001$), meaning that the two hair types have different rates of change across the length. The whisker base has significantly higher cuticle thickness than the body hair base ($p < 0.001$), potentially providing wear protection, as a cat's sparse facial whiskers contact external objects more frequently than its dense body hairs. However, the tips do not differ ($p = 0.737$) in cuticle thickness. This set of findings highlights distinct structural differences between body hairs and whiskers of the same animal; whiskers were previously shown to be thicker than body hair, but these two hair types were often assumed to have the same morphology~\cite{bryson_cortical_2009}. 

We analyzed the cortex composition from HAADF-STEM images of body hairs (\fig{fig:TEM_raw_whiskers}h) and whiskers (\fig{fig:TEM_raw_whiskers}i). The prevalence of granules in the body hair cortex increases by more than 200\% from the base ($4 \pm 0.15$\%) to the tip ($13 \pm 0.15$\%), meaning the cortex's compositional prevalence of keratin intermediate filaments decreases toward the tip. The composition of granules in cat whiskers exhibits the opposite trend. The whisker base resembles the body hair base in granule density; however, the granules entirely disappear at the whisker tip, and their compositional percentage is replaced by hollow porous zones with diameters of 200--500\,nm. Analyzing the cortex compositional percentage between hair types and locations (\fig{fig:TEM_raw_whiskers}j, \textcolor{teal}{Supplementary Table~1}) revealed significant effects of hair type ($F(1,38) = 67.05$, $p < 0.001$), location ($F(1,38) = 12.37$, $p = 0.001$), and their interaction ($F(1,38) = 60.34$, $p < 0.001$). Pure calcium~\cite{sun_mechanical_2015} is about 2000 times stiffer than pure keratin~\cite{wang_keratin_2016}, while melanin's modulus is on the order of megapascals~\cite{sarna2017nanomechanical}, closer to keratin than to calcium. The significant changes in granule prevalence from base to tip could cause mechanical differences for both body hairs and whiskers as melanin abundance has been shown to correlate with mechanical stiffness~\cite{sarna2019melanin}. Regions containing more granules or greater calcium enrichment would therefore be expected to be stiffer.

Mineral concentration also impacts the hardness and modulus of elasticity of mineralized or mineral-enriched biological materials~\cite{alkattan_influence_2018,ping_mineralization_2022}. We thus extracted the calcium-to-sulfur (Ca:S) intensity ratio from EDX spectra of the matrix and the CEGs; when no CEGs were present, the edges of voids were tested to allow comparisons across locations. The matrix has a very small Ca:S ratio of about $1.0\cdot 10^{-5}$ in all tested specimens. Analysis of these Ca:S ratios (\fig{fig:TEM_raw_whiskers}k--l) indicates that there are significant differences due to hair type ($F(1,16) = 309.31$, $p < 0.001$) and its interaction with tested location ($F(1,16) = 165.97$, $p < 0.001$). In other words, the Ca:S ratio of body hairs and whiskers changes differently according to the sampled location, but the effect of location itself is not quite significant ($F(1,16) = 3.98$, $p = 0.063$, \fig{fig:TEM_raw_whiskers}m). This finding aligns with the other cortex composition results (\fig{fig:TEM_raw_whiskers}j), as the bases of body hairs and whiskers seem similar, but their compositions diverge at the tips.

\subsection*{Hair stiffness and hardness gradients differ in cat body hair and whiskers}
\begin{figure}[thbp]
      \centering
      \includegraphics[width=1\textwidth,page=1,trim=0.0in 0in 0in 0.0in,clip]{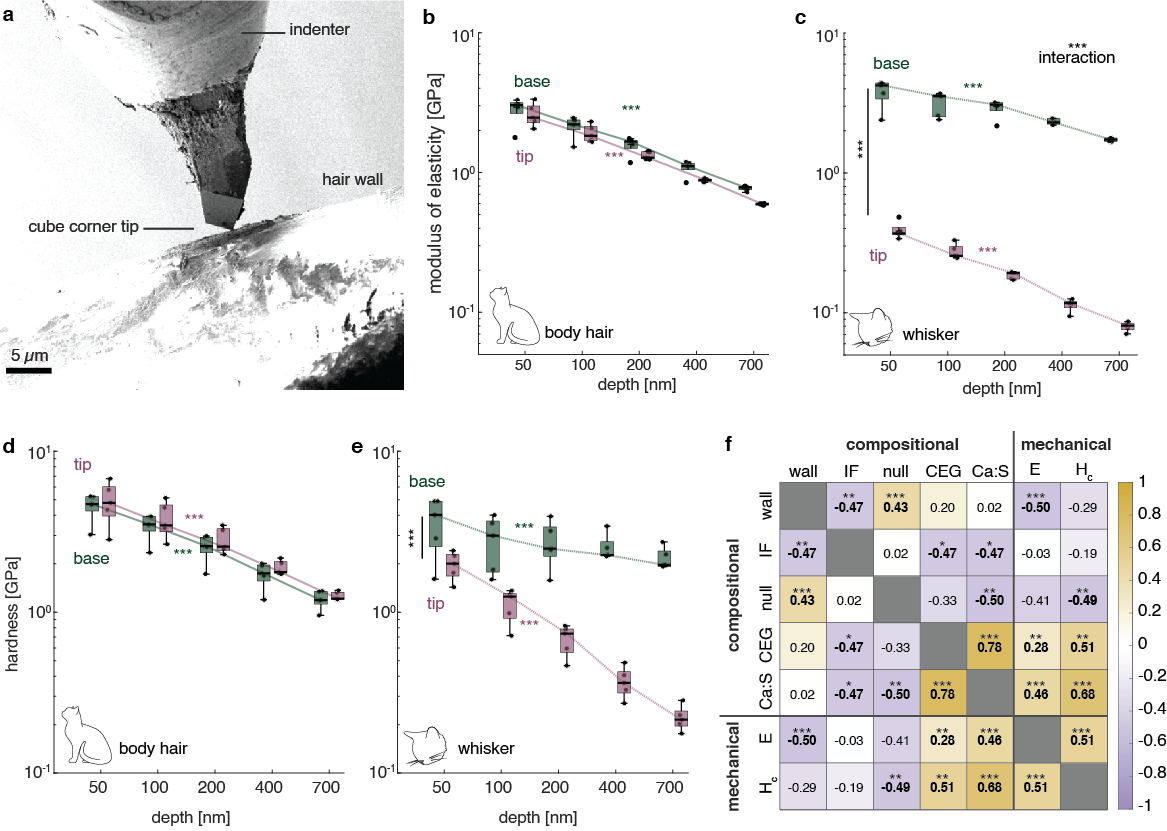}
      \caption{
      \textbf{$\mid$ Nanoindentation reveals that cat whiskers differ from body hairs in material stiffness and hardness.} \textbf{a,} Wide-angle SEM image of the indentation, where a boron-doped diamond cube-corner tip presses into the apex of the cuticle wall of a body hair. \textbf{b,} Modulus of elasticity ($E$) measured at a range of depths for the base ($n=5$) versus the tip ($n=5$) of cat body hairs. \textbf{c,} Modulus of elasticity of base ($n=5$) versus tip ($n=5$) of cat whiskers. \textbf{d,} Hardness ($H_c$) at a range of depths for the base ($n=5$) and tip ($n=5$) of cat body hair. \textbf{e,} Hardness of cat whiskers at the base ($n=5$) and tip ($n=5$). \textbf{f,} Pearson's correlation coefficient ($r$) calculated using bootstrapping between compositional and mechanical measurements of body hair base ($n=5$) and tip ($n=5$) and whisker base ($n=5$) and tip ($n=5$) using nanoindentation results from a depth of 50\,nm. Statistical significance is shown as: $^{\star}~p < 0.05, ^{\star\star}~p < 0.01, ^{\star\star\star}~p < 0.001$. For \textbf{b}--\textbf{e}, vertical lines indicate comparisons between base and tip, and green and purple stars indicate significant effects of depth for base and tip, respectively. For each plotted distribution, the line represents the median, and the dots show the raw data.
      }
      \label{fig:rawwhisker_indent}
\end{figure}

Nanomechanical testing elucidates the material properties of the specimens. Each intact hair segment is cyclically probed at the hair's apex using nanoindentation on the SEM stage (\fig{fig:rawwhisker_indent}a) with displacement control up to 1000\,nm. We analyzed the changes in modulus of elasticity ($E$, elastic recovery of the composite material) and hardness ($H_c$, plastic deformation of material contacted by the indenter) at five depths from 50 to 700\,nm into the body hair and whisker walls. During macroscopic testing, applied radial loads are supported by the entire hair; however, in our experiment, the nanoindenter's sharp tip accentuates localized material properties. Shallow nanoindentation thus mainly elucidates the properties of the cuticle wall, and deeper indentation increasingly involves the hair's cortex. Measurements are plotted logarithmically (\fig{fig:rawwhisker_indent}b-e) but analyzed linearly. 

Cat body hair modulus of elasticity (\fig{fig:rawwhisker_indent}b) is significantly affected by depth ($F(1.04,8.35) = 141.71$, $p < 0.001$), softening by about a factor of four from 50\,nm to 700\,nm. Cat body hair base ($2.82 \pm 0.54$\,GPa) is only 10\% stiffer than the tip ($2.61 \pm 0.44$\,GPa, $F(1,8)=2.24$, $p = 0.173$) at the 50\,nm depth, a non-significant difference. The effect of the indentation location (base versus tip) on the modulus of elasticity is more pronounced in the cat whisker ($F(1,8) = 533.45$, $p < 0.001$; \fig{fig:rawwhisker_indent}c), which at a 50\,nm depth has almost one order of magnitude reduction from the base ($3.28 \pm 0.75$\,GPa) to the tip ($0.38 \pm 0.05$\,GPa). A similar trend is observed across the five contact depths, with the depth also having a significant effect on the modulus of elasticity ($F(1.60,12.78) = 17.08$, $p < 0.001$), as well as their interaction ($F(1.60,12.78) = 9.27$, $p = 0.005$). The lowest modulus of elasticity across all sites was at the whisker tip, with the thickest cuticle and no CEGs present in the local cortex; the cuticle's main function does not seem to be adding mechanical stiffness to the hair. 

The hardness of cat body hair behaves similarly to its modulus of elasticity: hardness (\fig{fig:rawwhisker_indent}d) is significantly influenced only by depth ($F(1.05,8.36) = 89.77$, $p < 0.001$), with the outermost layers being the hardest. Cat whisker hardness (\fig{fig:rawwhisker_indent}e) also significantly decreases with indentation depth ($F(1.62,12.94) = 10.70$, $p = 0.003$). This measurement additionally shows a significant effect of location ($F(1,8) = 54.77$, $p < 0.001$), with the whisker base having higher hardness than the tip. Hardness changes of biomineralized materials are often attributed to the mineralized portions because the minerals provide additional resistance to plastic deformation~\cite{ping_mineralization_2022}. 

\subsection*{Granule abundance and calcium enrichment link compositional and mechanical properties}
After completing characterization, we explored potential trends between the compositional measurements determined through SEM, TEM, and EDX (cuticle wall thickness; area proportions of IF, null, and CEG; and Ca:S ratio) and the mechanical measurements determined through nanoindentation (elastic modulus and hardness at the shallowest depth of 50\,nm) for cat body hairs and whiskers (\fig{fig:rawwhisker_indent}f). By adapting the approach of Bala et al.~\cite{bala_respective_2011}, we identified several significant correlations between compositional and mechanical data across hair types and locations. Specifically, cuticle thickness has a significant moderate negative correlation with modulus of elasticity ($r = -0.51$, $p < 0.001$), indicating that our samples with thicker cuticles tend to have lower stiffness. Similarly, the percentage of hollow zones in the cortex has a significant moderate negative correlation with hardness ($r = -0.49$, $p = 0.006$), indicating more porous samples are less hard. Regarding granules, the correlations between granule prevalence and the mechanical parameters ($E$: $r = 0.28$, $p = 0.004$; $H_c$: $r = 0.51$, $p  = 0.004$) are significant but smaller than those between the mechanical parameters and calcium enrichment, as measured by the Ca:S ratio ($E$: $r = 0.46$, $p < 0.001$; $H_c$: $r = 0.68$, $p < 0.001$). Additional multivariate analysis (\textcolor{teal}{Supplementary Table~2}) revealed that CEG abundance ($\beta = 0.60$) and hardness ($\beta = 0.39$) are stable predictors of the Ca:S ratio, which could indicate that these three attributes of hair are closely related.

These correlations suggest that hair locations with a thicker cuticle wall (the tips) tend to have lower material stiffness and hardness than locations with thinner cuticle walls (the bases); this trend seems driven by the longitudinal softening of whiskers. Rather than influencing compliance, the granule-free cuticle could support the integrity of the hair through chemical and mechanical protection from the environment, consistent with previous studies~\cite{yang_strength_2020}. These same past studies have also shown that the cortex provides the mechanical structure of hairs by resisting mechanical deformations~\cite{yang_strength_2020}. Our analysis indicates that hair stiffness and hardness both positively correlate with the local granule abundance ($p<0.01$) and the ratio of calcium to sulfur in individual CEGs ($p<0.001$), providing evidence that compositional characteristics are linked to the mechanical properties of cat body hairs and whiskers. Since Ca:S ratio correlates with hardness across both hair types, we systematically apply Shindai chemical extraction --- a treatment that perturbs the chemical structure of the entire hair --- to investigate how this simulated environmental damage alters body hair structure, chemical composition, and mechanical properties.

\subsection*{Chemical extraction alters both the exterior and interior of body hair}
The cuticle wall is hypothesized to protect hair from liquids and chemical perturbations~\cite{yang_strength_2020,breakspear_cuticle_2022}. Since we found highly varying cuticle thickness in cat body hairs, with a thin wall at the base and a thick wall at the tip (\fig{fig:TEM_raw_whiskers}e), we investigated how chemical perturbations influence these hair structures both compositionally and mechanically. This systematic experiment chemically altered body hairs (both underfur and guard hairs) over distinct time intervals (0\,h, 72\,h, 120\,h) to create what we denote as treated hair (\textcolor{violet}{Extended Data Fig.~4}). Known as Shindai extraction~\cite{nakamura_rapid_2002}, the chemical etching technique that we employed is the standard for keratin extraction in hair and wool~\cite{deb2016isolation}; it allows analysis of both the chemically extracted solution and the solid extracted hair~\cite{brown_comparison_2016}. SEM imaging of treated body hair showed that its cuticle wall begins to delaminate after 72\,h (\fig{fig:extract_porosity}a--b). After 120\,h, the wall becomes smooth, as the rough cuticle scales have been etched away (\fig{fig:extract_porosity}c). SEM after etching for longer than 120\,h did not show further visual differences. 

To analyze how chemical extraction changes hair internally, we used micro-computed tomography (micro-CT) of cat body hairs that had been treated for increasing Shindai extraction times (0\,h, 72\,h, 120\,h, 168\,h, 336\,h). This imaging modality supports analysis of mineral density through intensity measurements~\cite{guntoro2019x}. The bases (\fig{fig:extract_porosity}d) and tips (\fig{fig:extract_porosity}d-e) were analyzed separately. The base structure experiences a large increase in mineral density after 72\,h with a large drop after 168\,h. The tip experiences a gradual increase until 120\,h, after which it declines similarly to the base. While micro-CT can give specific macro-based mineral information for biomineralized tissues, more precise measures are needed for calcium-enriched tissues like body hair.

\subsection*{Calcium abundance in solution surpasses keratin by four orders of magnitude}
\begin{figure}[thbp]
      \centering
      \includegraphics[width=1\textwidth,page=1,trim=0.0in 0in 0in 0.0in,clip]{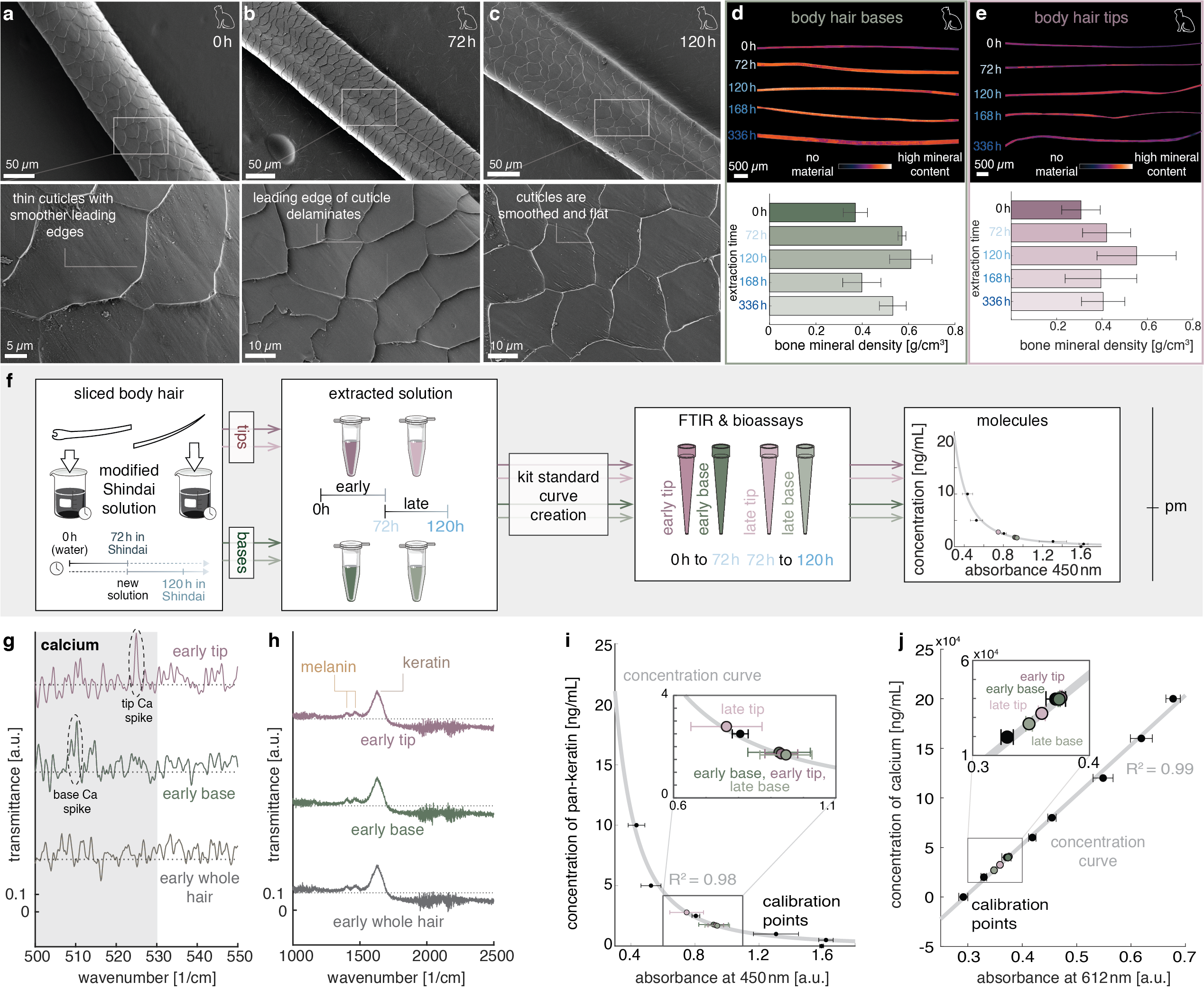}
      \caption{
      \textbf{$\mid$ Chemical extraction and assessment of keratin, melanin, and calcium in treated body hair.} \textbf{a,} SEM images of untreated body hair cuticle wall, showing curved scales with smooth leading edges. \textbf{b,} SEM images of body hair cuticle wall after 72\,h of Shindai extraction: the leading edges of the cuticles are delaminating from the wall. \textbf{c,} SEM images of body hair cuticle wall after 120\,h of Shindai extraction: the cuticle wall is smooth and flat, with no visible delamination. \textbf{d,} Micro-computed tomography (micro-CT) rendering to assess the bone mineral density ($n=3$ sets) of five body hair bases over Shindai treatment times. \textbf{e,} Micro-CT rendering to assess the bone mineral density ($n=3$ sets) of five body hair tips. \textbf{f,} Overview of separation of body hair bases and tips for selective early and late Shindai extraction; we aim to determine compositional differences during chemical extraction. \textbf{g,} Fourier transform infrared (FTIR) spectra of solution extracted from early (0--72\,h) treatment of whole hair, base, and tip, highlighting the wavenumbers where calcium transmits. \textbf{h,} FTIR spectra marked to show the melanin and keratin peaks found for early whole hair, base, and tip solutions. \textbf{i,} Extraction of pan-cytokeratin from the four tested solutions (\textcolor{violet}{Extended Data Fig.~9}a--d), including early ($n=3$) and late ($n=3$) base and early ($n=3$) and late ($n=3$) tip solutions, the calibration points ($n=21$), and the concentration curve's linear fit ($R^2 = 0.98$). \textbf{j,} Quantification of calcium ions in the four extracted solutions, with the calibration points ($n=21$) and the concentration curve's linear fit ($R^2 = 0.99$). 
      }
      \label{fig:extract_porosity} 
\end{figure}

As the base and tip responded differently to extraction after 72\,h, we performed selective extraction of body hair bases and tips; hairs were cut in half before treatment to allow preparation of three samples of equivalent mass (whole hair, bases, tips) (\fig{fig:extract_porosity}f). Although Shindai solution is the standard for extracting keratins~\cite{nakamura_rapid_2002,deb2016isolation}, the calcium content of the resulting solution was not previously assessed. Fourier transform infrared (FTIR) spectroscopy was thus applied to the solution produced after each sample had been extracted for 72\,h. FTIR shows that the base and tip solutions have different high peaks in the 500--530\,cm$^{-1}$ wavenumber range where calcium occurs, potentially showing different types of calcium organization (\fig{fig:extract_porosity}g). All three solutions show seemingly identical results for melanin and keratin content (\fig{fig:extract_porosity}h). This finding highlights that the Shindai technique extracts keratin from cat body hairs, as expected, and it simultaneously extracts melanin and calcium.

To further explore the chemical compositional differences along the hair, we performed pan-cytokeratin and calcium direct ion measurements on the 0--72\,h (early) Shindai solution and the 72--120\,h (late) Shindai solution for both hair bases and hair tips. Analysis of the cytokeratins reveals the concentration of 10 keratin genes that are typically found when whole hairs are extracted. Interestingly, for the early Shindai, both base and tip solutions exhibited the same pan-keratin yield. However, the late tip solution had double the keratin concentration compared to the early tip (\fig{fig:extract_porosity}i). The significantly thicker cuticle of the tip (\fig{fig:TEM_raw_whiskers}e,g) may help protect it from early extraction or environmental damage. 

Examination of the calcium concentration shows values between 2.5 and 4 $\times~10^4$\,ng/ml, which is about 10,000 times higher than the pan-keratin concentrations measured from the same solutions (\fig{fig:extract_porosity}j, \textcolor{teal}{Supplementary Table~3}). While researchers previously believed this method extracts only keratin from hair, our results clearly show that it also extracts a four-order-of-magnitude higher concentration of calcium (\fig{fig:extract_porosity}i--j): chemical extraction degrades CEGs and releases calcium-rich material into solution. This release may compromise the specificity of hair-based biomarker analyses, since extraction protocols are designed to target the keratin matrix~\cite{nakamura_rapid_2002}. Such interference could confound the quantification of cortisol concentration as a proxy for chronic stress~\cite{iob2019cardiovascular} or drug concentrations as a proxy for substance exposure~\cite{cuypers2018interpretation}. Having confirmed that the Shindai extraction technique chemically perturbs all components of the hair's composite structure (keratin structures, pigment granules, and elements), we now analyze how the compositional and mechanical properties of treated body hair change in response to these perturbations.

\subsection*{Chemical extraction differently perturbs cortex composition of hair bases and tips}
\begin{figure}[thbp]
      \centering
\includegraphics[width=1\textwidth,page=1,trim=0.0in 0in 0in 0.0in,clip]{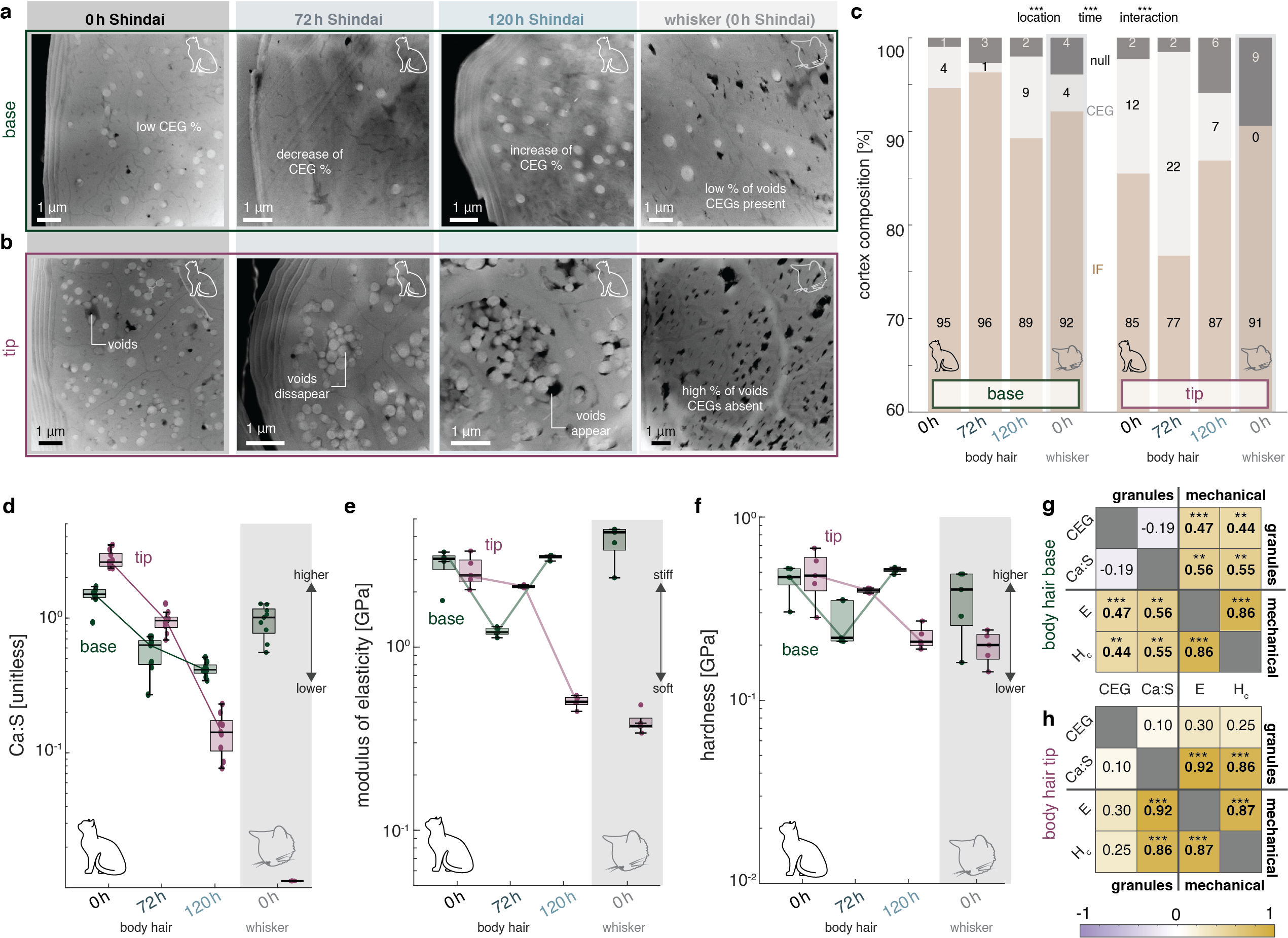}
      \caption{
      \textbf{$\mid$ Correlation of granule abundance and calcium enrichment with mechanical properties of chemically treated cat body hairs.} \textbf{a,} HAADF-STEM images of transverse sections of body hair base showing the compositional changes after 0, 72, and 120\,h of Shindai extraction, compared to an untreated whisker base. \textbf{b,} HAADF-STEM images of transverse sections of body hair tip after 0, 72, and 120\,h of Shindai extraction, compared to an untreated whisker tip. \textbf{c,} Compositional changes of 0, 72, and 120\,h chemically extracted base and tip shown as IF, granule (CEG), and null (hollow) mean percentages. All sample sites and extraction times have an equal number of measurements ($n=15$). \textbf{d,} Ca:S ratios of individual CEGs across the base and tip of body hairs for the different extraction times. All sample sites and extraction times have an equal number of measurements ($n=9$). \textbf{e,} Modulus of elasticity ($E$) from the indentation of the body hair base ($n=5$) and tip ($n=5$) across extraction times at the minimum contact depth (50\,nm). \textbf{f,} Hardness ($H_c$) from indentation of base ($n=5$) and tip ($n=5$) across extraction times at the minimum contact depth (50\,nm). In \textbf{a}--\textbf{f}, the same whisker measurements at the base (green) and tip (purple) are shown for comparison. For each distribution, the line represents the median and the dots the raw data. Statistics for the indentation of the extracted base and tip are included in \textcolor{violet}{Extended Data Fig.~6}a--d. \textbf{g,} Pearson's correlation coefficient ($r$) values of linear regression analysis for the body hair base across extraction times. \textbf{h,} Pearson's correlation coefficient ($r$) values of linear regression analysis for the body hair tip across extraction times. In \textbf{g}, \textbf{h}, statistical significance is shown as: $^{\star\star}~p < 0.01,\, ^{\star\star\star}~p < 0.001$. 
      }
      \label{fig:correlation}
\end{figure}

When analyzing cat body hair and whiskers together (\fig{fig:rawwhisker_indent}f), we found significant positive correlations between the mechanical properties of stiffness and hardness and the abundance of granules and the Ca:S ratio of CEGs (\textcolor{teal}{Supplementary Table~2}). Analysis of transverse STEM sections from the body hair base and tip after 0\,h (untreated), 72\,h, and 120\,h of extraction showed that the base (\fig{fig:correlation}a) and tip (\fig{fig:correlation}b) respond differently to chemical extraction. Analyzed ROIs of the cortex composition show that the prevalence of granules in the body hair base decreases after 72\,h and then increases after 120\,h of extraction (\fig{fig:correlation}c, \textcolor{teal}{Supplementary Table~4}): it is surprising that Shindai treatment changes the granule prevalence in both directions over time. The body hair tip shows an inverted evolution compared to the base: after 72\,h, there is a large increase in granule abundance. However, after 120\,h, the prevalence of granules in the body hair tip decreases dramatically to only 7\% of the cross-sectional area. Statistical analysis of the compositional changes Shindai extraction incurs in these body hairs (\fig{fig:correlation}c) shows a significant effect of the location ($F(1,28) = 108.57$, $p < 0.001$), time ($F(2,28) = 6.53$, $p = 0.003$), and their interaction ($F(2,28) = 56.62$, $p < 0.001$); specifically, the difference between tip and base is significant for 0\,h and 72\,h ($p < 0.001$), but not for 120\,h ($p = 0.388$). 

Proceeding to EDX analysis of individual CEGs seen in the STEM transverse sections, we find that the base's Ca:S ratio decreases significantly in the first extraction timestep (0 to 72\,h) and stays constant after 120\,h (\fig{fig:correlation}d). The tip experiences significant Ca:S changes over both extraction intervals, with the Ca:S ratio decreasing by a factor of about 20 from the untreated hair to the end of the experiment; the STEM images visually show this reduction as the CEGs take on a color similar to the matrix (\textcolor{violet}{Extended Data Fig.~5}) showcasing the granules are losing their calcium enrichment. Analysis shows the Ca:S ratio is significantly affected by the location ($F(1,16) = 122.32$, $p < 0.001$), time ($F(1.20,19.12) = 266.99$, $p < 0.001$), and their interaction ($F(1.20,19.12) = 49.44$, $p < 0.001$). As the chemical features have differing trends for the base and tip, we investigate the corresponding mechanical behavior for untreated and treated body hairs.

\subsection*{Prolonged chemical extraction stiffens hair bases and softens tips}
Indenting the body hair (underfur) base to 50\,nm (\fig{fig:correlation}e, \textcolor{violet}{Extended Data Fig.~6}a) shows that the elastic modulus declines from about 3\,GPa to about 1.2\,GPa after 72\,h and then increases after 120\,h back to the material stiffness of the untreated body hair base. Indentation of the tip (\textcolor{violet}{Extended Data Fig.~6}b) shows that 72\,h of extraction causes only a small decrease from about 2.5\,GPa to about 2.2\,GPa in elastic modulus; after 120\,h, the hair tips experience a sharp order-of-magnitude decline in modulus to 0.5\,GPa. Prolonged chemical extraction (72\,h to 120\,h) of the base and tip thus causes opposing effects: the hair base stiffens, and the tip softens. Furthermore, after 120\,h, we see that the body hair becomes almost identical to the modulus relationship of the untreated cat whisker (\fig{fig:correlation}e--f). Statistical analysis of the modulus of elasticity (\textcolor{violet}{Extended Data Fig.~6}a--b) shows significant effects of the location ($F(1,8) = 63.41$, $p < 0.001$), depth ($F(1.06,8.45) = 15.15$, $p = 0.004$), and time ($F(1.05,8.36) = 63.18$, $p < 0.001$). Bases and tips have significantly different elastic moduli, which significantly decrease as the indentation depth increases and vary significantly with extraction time. 

The hardness of the base and tip exhibits trends similar to the modulus of elasticity after 72\,h and 120\,h (\fig{fig:correlation}f). After 120\,h of Shindai extraction, the base becomes harder than 72\,h body hair ($p<0.001$), a common side effect of chemical over-treatment of human hair roots~\cite{wolfram_human_2003, miranda-vilela_overview_2014}. After 120\,h of Shindai, the hair tips become significantly softer, and the cuticle breaks down. This hyper-softening phenomenon may account for the development of split ends following extensive chemical treatment~\cite{taylor2024biomechanics}; the hair loses nearly all of the structural integrity characteristic of biopolymers, exhibiting behavior approaching that of an elastic material~\cite{lim_wanted_2019}. In statistically comparing the hardness data from base to tip (\fig{fig:correlation}f, \textcolor{violet}{Extended Data Fig.~6}c--g), we find there are significant effects of the location ($F(1,8) = 8.67$, $p = 0.019$), depth ($F(1.06,8.50) = 410.90$, $p < 0.001$), and time ($F(1.17,9.35) = 53.27$,  $p = 0.004$), as well as time $\times$ location ($F(1.17,9.35) = 40.28$,  $p < 0.001$), time $\times$ depth ($F(1.39,9.35) = 11.13$,  $p = 0.023$), and time $\times$ location $\times$ depth ($F(1.39,11.13) = 9.27$,  $p = 0.007$) interactions. The hardness values of bases and tips differ significantly; hardness significantly decreases with indentation depth and varies significantly with extraction time. 

To determine how these micro-mechanical structure differences impact macro-mechanical behavior, we performed uniaxial strain tests on raw, 72\,h, and 120\,h Shindai-extracted body hairs (\textcolor{violet}{Extended Data Fig.~7}a--b, \textcolor{violet}{Extended Data Table~1}). The measured macroscopic behavior follows the micro-structural changes of the base and tip. 72\,h of treatment causes the hairs to become more brittle, showing almost no plastic deformation (\textcolor{violet}{Extended Data Fig.~7}c), as well as stiffer (\textcolor{violet}{Extended Data Fig.~7}d). After 120\,h in solution, the hairs return to being more ductile and softer, similar to untreated hair. SEM images of the internal hair structures after failure show that untreated hair remains intact with small cracks appearing at the failure site (\textcolor{violet}{Extended Data Fig.~8}a--b). By contrast, for hair that has been treated for 72\,h, the cuticle completely unzips, showing internal delamination of keratins (\textcolor{violet}{Extended Data Fig.~8}c--d). 

\subsection*{Granule calcium content matches mechanical behavior of untreated and treated bases and tips}
To further investigate the correlation between compositional parameters (granule abundance and Ca:S ratio in the CEGs) and mechanical properties (modulus of elasticity and hardness at 50\,nm depth), we evaluate them separately for the hair bases and tips across treatment times. Analysis of the treated body hair bases (\fig{fig:correlation}g, \textcolor{teal}{Supplementary Table~5}) shows significant correlations across all four granule-mechanical connections ($0.47 < r < 0.92$), indicating that both the abundance of calcium-enriched granules and the ratio between calcium and sulfur correlate strongly with the hair mechanics. As our experiments have indicated, prolonged treatment (120\,h Shindai extraction) leads to stiffening and hardening of the hair base. 

Proceeding to the body hair tips, we find that the Ca:S ratio has a very strong positive correlation with the modulus of elasticity ($r = 0.92$, $p < 0.001$) and a strong positive correlation with the hardness ($r = 0.88$, $p < 0.001$) (\fig{fig:correlation}h), indicating that the prevalence of calcium in the granules may determine the tip's compliance. Indeed, multivariate analysis (\textcolor{teal}{Supplementary Table 6}) suggests that the Ca:S ratio in the tip may be predicted by the modulus of elasticity. Prolonged treatment (120\,h Shindai extraction) softens and reduces the hardness of the tip; treated body hair tips have no CEGs (\fig{fig:correlation}c), are very soft (\fig{fig:correlation}e), and contain many voids (\fig{fig:correlation}b). This finding, combined with the macro-mechanical analysis of over-extracted hair, shows that chemically treated hair tips may split because of their highly porous and soft structure.

\subsection*{Conclusion}
Cat body hairs and whiskers exhibit three structural characteristics at distinct length scales that combine to produce biological composites with longitudinally varying material properties. First, the layered material barrier of the granule-free \textit{cuticle wall} defends the granule inclusions and calcium enrichment located in the inner cortex against chemical extraction. In contrast to prior assumptions of constant thickness, the cuticle wall significantly increases in thickness from the base to the tip of both body hairs and whiskers. This finding is surprising because the hair tip grows first, and it seems unlikely that hair grows with a progressively thinning cuticle from tip to base. Another possible explanation for this pattern is that cuticle layers could form passively through environmental oxidation. Contact with oxygen could destabilize keratin's alpha-helix by breaking down disulfide bonds~\cite{perta2021closing}, causing alpha-keratins to transition to beta-keratins in the cuticle~\cite{stanic2015local}, with longer exposure producing additional layers or splitting layers (\fig{fig:TEM_raw_whiskers}e). A thicker cuticle wall likely also provides better wear resistance for exposed hair tips.

Second, at a smaller scale, we discovered that the 250-nm-diameter melanosome-like granules in the cortex are calcium enriched, longitudinally aligned, and variable in abundance from base to tip. The presence of stiffer oblong structures longitudinally arranged between the softer keratin intermediate filaments seems to stiffen the entire composite structure using a mechanism potentially similar to fiber jamming~\cite{Aktas21-AFM-Jamming}. Body hairs have significantly higher granule abundance at the tip compared to the base. Cat whiskers have a complete absence of granules or calcium enrichment at their elastomeric tips, potentially creating the functionally graded stiffness that has been hypothesized to encode contact along the whisker's length~\cite{schulz2026functional}. Third, the \textit{presence of calcium} in individual calcium-enriched granules increases from the base to the tip of body hairs and can be selectively chemically extracted using the Shindai approach. The extracted calcium concentration is 10,000 times higher than that of the extracted cytokeratins across all extraction trials of both hair bases and tips and shorter and longer extraction times, underlining the high concentration of calcium throughout the entire hair length. Furthermore, this chemical extraction of the hair's composite structure significantly changes the mechanical properties of hair, ending with stiff bases and soft, split tips.

The hierarchy of these three structural characteristics demonstrates that the body hairs and whiskers of domestic cat are complex, functionally graded biological composites. The local intensity of calcium enrichment correlates moderately with mechanical properties consistent with those of biomineralized tissues. Based on these species-specific findings, we propose that the keratin-based cortex of mammalian hairs warrants broader investigation to determine whether the combination of granule composition and calcium enrichment modulates hair mechanics across species. Future experiments targeting specific granule or calcium perturbations are required to determine whether CEGs directly control stiffness or act together with the local arrangement of the keratin intermediate fibers. Beyond possible implications for healthcare and personal hygiene products, these discoveries could inspire methods for selectively extracting and generating new types of materials, potentially allowing more precise recycling of widely available keratin-based biomaterials like hair~\cite{wang_entropy-driven_2025}. 

\subsection*{Acknowledgments}
The authors thank R. Henderson for discussions on protein structural chemistry, L. Ting for mechanics discussions, B. Akta\c{s} for discussions of jammed biological structures, C. Higgins for discussions of hair mechanics, P. Yunker and W. Ratcliff for biophysics discussions, T. Kucheniene and R. Griego for assistance with sample acquisition, K. Angarra for conversations on keratin, A. Southan for biopolymer discussions, H. David for help with SEM images, O. Mohamed for bioassay analysis, K. Araslanova and E. Maru for data visualization assistance, P. Karvelis for the \textit{daviolinplot} repository, J. Fiene for manuscript and title input, and L. Wang for optical microscopy help. This work was supported by the Alexander von Humboldt Foundation (to A.K.S.), the Carl Schneider Stiftung (to G.R. and K.J.K.), and the Max Planck Society (to J.L., P.A.v.A., G.R., and K.J.K.).

\subsubsection*{Author Contributions} A.K.S., G.R., and K.J.K. conceived the idea and designed the experiments. A.K.S., P.A.v.A., G.R., and K.J.K. supervised the research project. A.K.S., S.G., V.S., N.G.-V., D.S., F.P., C.B.D., D.S.P., H.V., J.L., and G.R. conducted the experiments. A.K.S., S.G., V.S., G.B., N.G.-V., D.S., F.P., C.B.D., D.S.P., J.L., P.A.v.A., G.R., and K.J.K. analyzed the data. All authors contributed to the writing and editing of the manuscript. 

\subsubsection*{Ethics declarations}
The authors declare no competing interests. 

\subsubsection*{Data and materials availability:} 
All data and statistical results are available in the main text, supplementary materials, or Edmond online repository~\cite{schulz_dataset_2026}. Code for the general linear models (GLMs), correlation analysis, visualization, and multivariate analysis is available from the following GitHub repository: \href{https://github.com/GiuliaBallardini/MorphCompMech}{https://github.com/GiuliaBallardini/MorphCompMech}. 

\clearpage
\clearpage

\clearpage

\clearpage

\clearpage

\clearpage

\singlespacing
\scriptsize
\begin{footnotesize}

\end{footnotesize}
\bibliographystyle{naturemag}

\end{document}